\documentstyle[11pt]{article}
\newcommand {\prz}{\partial}
\newcommand {\bq}{{\bf q}}
\newcommand {\be}{\begin{equation}}
\newcommand {\ee}{\end{equation}}

\newcommand {\vth}{\vartheta}

\begin{document}
\LARGE
\begin{center}
A model of dispersion in the unsteady separated shear flow past complex geometries \small \\
\bigskip
P. FRANZESE$^1$ AND L. ZANNETTI$^2$\\
\smallskip
\scriptsize\rm
$^1$Institute for Computational Sciences and Informatics, George Mason University, Fairfax, Virginia 22030, U.S.A.
$^{2}$Dipartimento di Ingegneria Aeronautica, Politecnico di Torino, 10129 Torino, Italy
\end{center}

\renewcommand{\baselinestretch}{1.5}    
\small

\section*{Abstract}

Separated flows past complex geometries are modelled by discrete
vortex techniques. The flows are assumed to be rotational and inviscid, and a new technique is described to determine the streamfunctions for linear shear profiles. The geometries considered are the snow cornice and the backward-facing step, whose edges allow for the separation of the flow and reattachment downstream of the recirculation regions. A point vortex has been added to the flows in order to constrain the separation points to be located at the edges, while the conformal mappings have been modified in order to smooth the sharp edges and let the separation points be free to oscillate around the points of maximum curvature. Unsteadiness is imposed on the flow by perturbing the vortex location, either by displacing the vortex from equilibrium, or by imposing a random perturbation with zero mean on the vortex in equilibrium. The trajectories of passive scalars continuously released upwind of the separation point and trapped by the recirculating bubble are numerically integrated, and concentration time series are calculated at fixed locations downwind of the reattachment points. This model proves to be capable of reproducing the trapping and intermittent release of scalars, in agreement with the simulation of the flow past a snow cornice performed by a discrete multi-vortex model, as well as with direct numerical simulations of the flow past a backward-facing step. The simulation results indicate that for flows undergoing separation and reattachment the unsteadiness of the recirculating bubble is the main mechanism responsible for the intense large-scale concentration fluctuations downstream.

\section{Introduction}

Transport and dispersion of passive scalars in turbulent boundary layers over complex geometries is an important problem of practical relevance, which has been widely investigated in recent years mainly because of a growing concern about the environmental issue, but also because of its engineering applications such as optimization of combustors or evaluation of mixing properties of flows in chemical reactors.  

However, the complexity of flows in which separation and reattachment occur make them less suitable to analytical studies than regular boundary layer flows. Relatively little is known about the behaviour of a plume near hills, or buildings, or, in general, over irregular terrain (Hosker 1984). For a classical problem like the flow over a backward-facing step, there is not an applicable mathematical technique for modeling the evolution of the scalar field. Laboratory experiments generally provide an Eulerian portrait of this flow (e.g. Bradshaw \& Wong 1972), but with no insight into the dispersion process. More recently, Le, Moin \& Kim (1997) performed a direct numerical simulation of a backward-facing step for Reynolds numbers as high as 5100 revealing, among the others, some of the unsteady flow characteristics. They detected the oscillation of the (spanwise-averaged) reattachment location, in agreement with the results of several other works (e.g. Driver, Seegmiller \& Marvin 1983, 1987). The shear layer composed of many small high-intensity vortices, which extends to the reattachment point, rolls up forming a large-scale structure in the recirculation region: the periodic detachment of this large-scale vortex from the step causes the oscillation of the reattachment point.
The main aim of this paper is to model the unsteadiness of reattached flows, and to assess how it affects the dispersion of passive scalars released by a continuous point source. 

We will model separated flows by means of point vortex techniques,
assuming the flow as two-dimensional and inviscid. We will consider two different geometries: a backward-facing step, and a ``snow cornice''. Snow cornices are natural devices that control the flow separation on mountain crests by trapping vortices. The modeling technique is not new: Ringleb (1961) studied the steady separated flow past a snow cornice assuming a two dimensional potential flow, which allowed him to solve the problem analytically by means of the classical method of conformal mapping. To represent a snow cornice, Ringleb devised mappings to transform the real axis in the complex $\zeta$-plane onto a line forming a sharp edge in the complex $z$-plane. In this way the complex potential of a uniform flow over the real axis in the $\zeta$-plane was ascribed to the flow on the corresponding region in the $z$-plane. A point vortex (together with its reflected image) was added to the flow to model the recirculating region. The vortex was in equilibrium and satisfied a steady Kutta condition. In a more recent paper, Cortelezzi, Leonard \& Doyle (1994) apply point vortex techniques to model the unsteady separated flow over a semi-infinite plate, and to assess active control techniques. 

Because of the incompressibility condition, the Lagrangian motion of a passive scalar is governed by a Hamiltonian system whose conjugate variables are the Cartesian coordinates of the scalar, and whose Hamilton function is the streamfunction. The motion of a point vortex is governed by a Hamiltonian system as well, where the Hamilton function is related to the Green function for the Laplacian of the flow region (Masotti 1931; Lin 1941). 

In this respect, the motion of the vortex in the Ringleb model is
integrable since it is governed by a one-degree of freedom autonomous
Hamiltonian system. Incidentally, although we found that Ringleb made
a formal error in determining the law of the motion of a vortex in
the presence of a wall, and therefore failed to derive the correct
equilibrium conditions, his results are qualitatively correct (see Appendix B).

The streamfunction, namely the Hamiltonian of the fluid particles,  depends on the vortex position and therefore is in general time-dependent (non-autonomous). For such a system we expect a non integrable, chaotic, particle motion (Novikov \& Sedov 1979; Aref 1983). However, if the vortex is located at its equilibrium point, the Hamiltonian of the particles is autonomous and the motion is integrable and regular. For the integrable case, the flow field is virtually divided in two main bodies: the fluid entrained by the vortex (closed streamlines encircling the vortex), and the free-stream flow (open streamlines). The two regions do not exchange fluid with each other and are separated by a streamline which leaves the sharp edge of the wall, and reattaches downstream (i.e. a heteroclinic orbit). For instance, in the transformed $\zeta$-plane this flow is a vortex pair, where the real axis is the line of symmetry.

The present model is different from Ringleb's in three main aspects: 

(i) We assume a shear flow as the asymptotic upstream condition in order to have a more meaningful flow than the potential one, but still simple enough to allow for an analytical study. 

(ii) The system has been made unsteady by perturbing the equilibrium of the vortex. 

(iii) Finally, we consider geometries where the sharp corners are
smoothed to blunted edges with large but finite curvature, to satisfy
an unsteady Kutta condition for a vortex of constant circulation not
in equilibrium.   

The large curvature edge still causes separation, in that according to the Kutta condition it is possible to define a vortex whose strength produces separation. A vortex slightly displaced from its stable equilibrium moves along a small periodic orbit, and causes the oscillation of the detachment point around the maximum curvature point. Since the edge is smooth, the flow around the edge does not have any singularities and the Kutta condition can be considered fulfilled.

A typical representation of the mean flow vertical profile by power laws is
used in the atmospheric boundary layer in stable conditions. For instance, a linear shear profile has been considered by Hunt, Leibovich \& Richards (1988) to study how the shear affects the flow past hills with low slopes.

On the transformed $\zeta$-plane, the flow reduces to the oscillating vortex pair (OVP). Rom-Kedar, Leonard \& Wiggins (1990) studied the OVP flow extensively, and elucidated the mechanisms of the phase-space Lagrangian chaos. The stable and unstable manifolds of the heteroclinic orbit intersect each other, and drive the fluid to be entrained and detrained by the vortex. The Eulerian counterpart of this phenomenon is the chaotic mixing of the fluid in the region surrounding the separatrix streamline. We are interested in the effects of these phenomena on the dispersion of scalars.  

We investigated how the separation and reattachment of an unsteady flow influence the behaviour of effluent plumes released near or inside the recirculating region. We found that the presence of the recirculation bubble dramatically affects the large-scale dynamics of passive scalars, causing important fluctuations of the concentration field downstream.

In the following sections the Ringleb model for steady potential flow is recalled, then a method is proposed to study analytically unsteady rotational flows with constant vorticity. Dynamical systems formalism is used to find equilibrium configurations, to discuss their stability and to recognize and quantify chaotic mixing. Numerical simulations for two different geometries are presented, along with the concentration time series at several sampling points.

\section{Mathematical formulation of the steady potential flow}

In this section we describe a mathematical model of recirculating
flow over complex geometries such as a snow cornice and a backward-facing step. This model, which was proposed by Ringleb (1961), is able to represent a two-dimensional steady separation from the edge. It constitutes the basis for the model of unsteady, recirculating shear flow, which we present in \S 3.  

We consider a region bounded by a piecewise analytical curve in the $z$-plane, where $z=x+iy$. The solid boundary extends to infinity along the $x$-axis, and is characterized by a sharp corner. As a consequence of the Riemann mapping theorem (see, e.g., Nehari 1975; Henrici 1974), such a region can be obtained as the conformal image of the half-plane $\eta\ge 0$ in the $\zeta$-plane, where $\zeta=\xi+i\eta$. The infinity in the $z$-plane corresponds to the infinity in the $\zeta$-plane. For instance, the function suggested by Ringleb (1961), 
\be
z=\zeta+\frac{\zeta_1^2}{\zeta-\zeta_1}\label{a1}
\ee
where $\zeta_1=\xi_1+i\eta_1$ is a complex constant with $\eta_1<0$, yields to the geometry plotted in figure 1a. The backward-facing step shown in figure 1b can be obtained by the Schwartz-Christoffel mapping:
\be
z=\sqrt{\zeta^2-1}+\log \left(\zeta+\sqrt{\zeta^2-1}\right)\;.\label{a2}
\ee


The model consists of a vortex of constant strength which satisfies a Kutta condition, in equilibrium with a steady free-stream velocity.
The complex potential of the flow is built by superimposition of basic flows: in the $\zeta$-plane it is expressed by the function:
\be
w=q_\infty\zeta+\frac{\gamma}{2\pi i}\log\left(\frac{\zeta-\zeta_0}{\zeta-\zeta_0^\star}\right)\label{a3}
\ee
where $q_\infty$ is the free-stream velocity, and the second term on the right hand side is the complex potential of a point vortex of strength $\gamma$ located at $\zeta_0$ and of its image at the complex conjugate position $\zeta_0^\star$. Equation (\ref{a3}) represents the complex potential of a vortex pair of opposite signs in the $\zeta$-plane (see figure 7). 

Since the complex potential is invariant under a conformal mapping $z=z(\zeta)$, the constant $q_\infty$ is equal to the free-stream velocity $Q_\infty$ in the $z$-plane:  
\be
Q_\infty=\lim_{z\rightarrow\infty} \frac{dw}{dz}=\lim_{\zeta\rightarrow\infty}\frac{dw}{d\zeta} \left(\frac{dz}{d\zeta}\right)^{-1}=q_\infty\;.
\label{a4}
\ee

The trajectories of a vortex in the physical $z$-plane are the solutions of the Hamiltonian system:
\be
\dot x_0=\frac{\prz H}{\prz y_0};\;\;\;\;\;\;\;\;\;\;\;\;\;\;\; \dot y_0=-\frac{\prz H}{\prz x_0}\label{a6}
\ee
where the Hamiltonian $H$ can be derived from the Hamiltonian $H'$ of a vortex in the $\zeta$-plane, according to the Routh rule (e.g. Clements 1973):
\be
H=H'+\frac{\gamma}{4\pi}\log\left|\frac{d z_0}{d \zeta_0}\right|\label{a7}
\ee
where:
\be
H'=q_\infty\eta_0+\frac{\gamma}{4\pi}\log{\eta_0}\;.
\label{a8c}
\ee
Derivations of the Hamiltonian $H'$ and the Routh transformation
rule are given in Appendix A.
\subsection{Steady Kutta condition}
The separation of a flow around a sharp or a large curvature corner is captured by the model if it satisfies the Kutta condition, that is, a relationship between the strength of the vortex and its position in order to eliminate the singularity of the flow. Therefore, the condition of zero velocity at the edge is imposed:
\be
\left(\frac{dw}{d\zeta}\right)_{\zeta=\zeta_c}=0
\label{b1}
\ee
where $\zeta_c$ indicates the location of the corner in the $\zeta$-plane. 

For the complex potential $w(\zeta)$  (\ref{a3}), the Kutta condition is satisfied when
\be
\frac{\gamma}{q_\infty}=-\frac{\pi|\zeta_c-\zeta_0|^2}{\eta_0}
\label{b2}
\ee
where we notice that there are infinite possible values for the ratio $\gamma/q_\infty$. For the mapping (\ref{a1}) $\zeta_c=0$, whereas $\zeta_c=-1$ for (\ref{a2}). 

On the physical plane, the complex velocity at the corner $z_c$ is
\be
\lim_{z\rightarrow z_c}\frac{dw}{dz}=\lim_{\zeta\rightarrow \zeta_c}\frac{dw}{d\zeta}\frac{d\zeta}{dz}
\label{b3}
\ee
whose value is finite when the corner is a cusp, as in the case of mapping (\ref{a1}), while it is null in the other case.   

In order to complete the modeling of the recirculating flow, we are
now concerned with finding the equilibrium conditions for the vortex. We recall that the phase space of the dynamical systems (\ref{a6}) coincides with the flow region, in the sense that the Cartesian coordinates of the vortex ($x_0$, $y_0$) represent the conjugate variables, and the level lines of the Hamiltonian $H$ form the pattern of the possible trajectories of a vortex of assigned strength. Therefore the equilibrium locations ($\dot x_0=0$, $\dot y_0=0$) are the fixed points of this map: the elliptic points correspond to stable equilibrium, the hyperbolic ones to unstable equilibrium. The model of a steady recirculating flow behind a corner is represented by a vortex in stable equilibrium (i.e.  a {\em trapped vortex}), and satisfying the Kutta condition. 
The solutions to the equilibrium equations, if any, depend on the geometry of the solid boundary and on the ratio of the vortex intensity to the asymptotic velocity, i.e. $\gamma/q_\infty$. Since the geometries we considered were found to allow vortex capturing, we expect a set of stable solutions.

Therefore, the vortex coordinates ($\xi_0,\eta_0$) on the $\zeta$-plane have to satisfy simultaneously the equilibrium condition and the Kutta condition (\ref{b2}). They are the solutions of the equation:
\be
\frac{|\zeta_c-\zeta_0|^2}{4\eta_0}\left\lbrack\frac{1}{\eta_0}\frac{d\zeta_0}{dz_0}-i\frac{d}{dz}\left(\log
\frac{d\zeta_0}{dz_0}\right)\right\rbrack=1\;.
\label{b4}
\ee
The vortex strength is then obtained by the Kutta condition (\ref{b2}).
\subsection{Comparison with a multiple vortex method}
The accuracy of the above vortex model has been checked by comparing some of the results with those generated by a more sophisticated vortex method. We used a discrete vortex method (Ceschini 1993; Ferlauto 1996) to describe the impulsive start of the flow characterized by vortex shedding. We performed the simulation for the snow-cornice geometry. The method consists of assuming the initial flow around the wall as potential, and satisfying an unsteady Kutta condition by adding a free point vortex of suitable strength in a prescribed location close to the corner. The vortex moves because of, and according to, the wall and the background flow. The motion is then numerically integrated. At given time intervals a new vortex is introduced into the flow to restore the regularity. Figure 2 shows the pattern of the vortices shed throughout the transient, where the vortices strengths are proportional to the symbol sizes. The transient is characterized by a vortex sheet rolling up around the point of highest absolute circulation. This result is consistent with the description of the transient as given by Cortelezzi, Leonard \& Doyle (1994) and Le, Moin \& Kim (1997) for other vortex trapping flows. The vortex sheet increases in size and circulation until eventually reaching a more steady configuration, with small oscillation of its global circulation and centroid.

We compared the asymptotic circulation and the centroid location with
the strength and location of the single point vortex used in our
model (the centroid location is evaluated assuming the strengths as
masses). We plot in figure 3a the total circulation as a function of
time (solid line), as well as the circulation for the single vortex
model (dotted line). In figure 3b we plot the distance of the centroid for
the multiple vortex method from the equilibrium point of a single
vortex. The two models are in good agreement with each other as far
as the intensity of the vorticity field and the location of the
centre of vorticity are concerned, showing that the large-scale characteristics of the flow are well captured by a single vortex model. Of course, the strength of the single vortex model is in its simplicity, which permits the analytical description of a large-scale mechanism of dispersion, as discussed in \S 4.




\section{Shear flow}
In this section we describe a method to model the separation and recirculation of a shear flow over complex geometries. 

We assume a rotational flow field, with vorticity $\nabla \times \bq=\omega=const$ and with the following velocity profile in the far field:
\be
\lim_{x\rightarrow\pm\infty}q=u_\infty-\omega y\;.\label{c1}
\ee
The motion of passive scalars is governed by the Hamiltonian dynamical system:
\be
\dot x=\frac{\prz \psi}{\prz y};\;\;\;\;\;\;\;\;\;\;\;\;\;\;\; \dot y=-\frac{\prz \psi}{\prz x}\label{a5}
\ee
where the streamfunction $\psi$ represents the Hamilton function and the scalar coordinates ($x,y$) are canonical variables. 
The streamfunction $\psi$ is related to the vorticity $\omega$ by the linear Poisson equation:
\be
\nabla^2\psi=-\omega\;.\label{c2}
\ee
To solve the Poisson equation (\ref{c2}) we developed an analytical procedure based on conformal mapping, which is inspired by the method described by Tsien (1943).\\
Let us assume
\be
\psi=\psi_\omega+\psi_p\label{c2a}
\ee
where the streamfunction $\psi_\omega$ is relevant to the shear flow $u=-\omega y$, i.e.
\be
\psi_\omega=-\frac{1}{2}\omega\; y^2\label{c3}
\ee
and the function $\psi_p$ is harmonic in the flow region.\\
Since
\be
\nabla^2\psi_\omega=-\omega\label{c4}
\ee
and
\be
\nabla^2\psi_p=0\label{c5}
\ee
equation (\ref{c2}) is satisfied. However, the problem has been
reduced to the determination of the harmonic streamfunction $\psi_p$
which ensures the fulfillment of the boundary conditions, that is,
the far field velocity equal to $(u_\infty-\omega\,y)$ and the
impermeability of the wall.     

Therefore, we seek a complex potential $w_p$ whose imaginary part is
$\psi_p$. We define the flow regions in the $z$-plane as the conformal representation of the half-plane $\Im(\zeta)>0$ according to the mappings:
\be
z=\zeta+\frac{\zeta_1^2}{\zeta+i\delta -\zeta_1}\label{c6}
\ee
and
\be
z=\sqrt{(\zeta+i\delta)^2-1}+\log\left(\zeta+i\delta+\sqrt{(\zeta+i\delta)^2-1}\right)\label{c7}
\ee
which are the modified Ringleb mapping (\ref{a1}) and the Schwartz-Cristoffel mapping (\ref{a2}), respectively, where we introduced the real positive parameter $\delta$ in order to smooth the corners (the smaller $\delta$, the sharper the corner).

The further mapping:
\be
\lambda=\frac{i-\zeta}{i+\zeta}\label{c7a}
\ee
transforms the half-plane $\Im(\zeta)>0$ onto the interior of the
unit circle $|\lambda|=1$ in the $\lambda$-plane. Summing up, the chain mapping $z\rightarrow\zeta\rightarrow\lambda$ maps the physical boundary of the flow region onto the unit circle in the $\lambda$-plane, where $\lambda=-1$ corresponds to infinity in the $z$-plane.

On the physical plane, the shear flow (\ref{c3}) induces the following velocity component $\tilde{v}$ normal to the boundary:
\be
\tilde{v}=\omega\; y_b\; \sin \beta\label{c7b}
\ee
where the subscript $b$ denotes boundary, and $\beta=\arg\left(dz/d\zeta\right)$ is the angle between the boundary and the $x$-axis.

According to equation (\ref{c7b}), the condition of impermeability for the boundary in the $z$-plane is satisfied when the velocity component normal to the boundary, due to the complex potential $w_p$, is equal to $-\tilde{v}$, that is:
\be
-\Im\left(\frac{dw_p}{dz}e^{i\beta}\right)=-\tilde{v}\label{c9}
\ee
which translates on the $\lambda$-plane as
\be
\Re\left(\frac{dw_p}{d\lambda}\lambda\right)_{|\lambda|=1}=-\tilde{v}\left|\frac{dz}{d\lambda}\right|
_{|\lambda|=1}\;.
\label{c10}
\ee
 
As there are no point vortices (they can be added to the flow later), the complex velocity $dw_p/dz$ cannot be singular inside the flow field. Since the mappings $z\rightarrow\lambda$ are regular inside the unit circle $|\lambda|\le 1$ ($\lambda\ne -1$) in the $\lambda$-plane, then the complex function $dw_p/d\lambda$ must not be singular inside the unit circle as well, except for the point $\lambda=-1$ that corresponds to $z=\infty$. Therefore we can assume:
\be
\frac{dw_p}{d\lambda}=F(\lambda)+\sum_{n=1}^\infty(a_n-ib_n)\lambda^{n-1}\label{c11}
\ee 
where $F(\lambda)$ contains the possible singularities located in $\lambda=-1$, that is, in general,
\be
F(\lambda)=\sum_{j=1}^\infty c_j (\lambda+1)^{-j}\;.
\label{c12}
\ee 

To determine the $c_j$ coefficients, we recall that the velocity at infinity ($z=\infty$), due to the complex potential $w_p$, is $u_\infty$. Since for the mappings (\ref{c6}) or (\ref{c7}) the derivative at infinity is
\be
\lim_{z\rightarrow\infty}\frac{d\zeta}{dz}=1\label{c13}
\ee
we have
\be 
u_\infty=\lim_{\zeta\rightarrow\infty}\frac{dw_p}{d\zeta}=
\lim_{\lambda\rightarrow -1}\left(\frac{dw_p}{d\lambda}\frac{d\lambda}{d\zeta}\right)
=\lim_{\lambda\rightarrow -1}\left[\frac{dw_p}{d\lambda}\;\frac{(\lambda+1)^2}{-2i}\right]
\label{c14}
\ee
which implies that all the coefficients $c_j$ for $j>2$ are null, and that
\be
c_2=-2i\;u_\infty\;.
\label{c15a}
\ee

We determine the coefficient $c_1$ by considering that the mass flow $Q$ across the boundary, due to the shear flow, has to be balanced by the potential flow. We have
\be
Q=\int_{-\infty}^{+\infty}-\omega\; y_b\;\frac{dy_b}{dx_b}\;dx_b=-\frac{\omega}{2}\;\left[ y_b^2\right]_{-\infty}^{+\infty}
\label{c15}
\ee
and also, according to the invariance property of the complex potential,
\be
-Q=-i\;\oint_{|\lambda|=1} \frac{dw_p}{d\lambda}= \pi\;c_1\;.\label{c16}
\ee 
The function $F(\lambda)$ is then fully determined:
\be
F(\lambda)=-2i\;u_\infty\;\frac{1}{(\lambda+1)^2}-\frac{Q}{\pi}\;\frac{1}{\lambda+1}
\label{c17}
\ee
According to equation (\ref{c15}), the mass flow $Q$ relevant to the mapping (\ref{c6}) is null, while for the mapping (\ref{c7}) it is
\[Q=-\frac{\omega}{2}(\pi^2+2\pi\;\delta)\;.\]

To calculate the second term on the right hand side of equation (\ref{c11}), we combine equation (\ref{c11}) with equations (\ref{c10}) and (\ref{c17}) evaluated at $|\lambda|=1$, that is $\lambda=\exp(i\vth)$: 
\begin{eqnarray}
\sum_{n=1}^\infty(a_n\cos n\vth+b_n\sin n\vth)&=&-\tilde{v}\left|\frac{dz}{d\lambda}\right|
-\Re\left(-2i\;u_\infty\;\frac{\lambda}{(\lambda+1)^2}-\frac{Q}{\pi}\;\frac{\lambda}{\lambda+1}\right)\nonumber\\
 &=& -\tilde{v}\left|\frac{dz}{d\lambda}\right|+\frac{Q}{2\pi}\;.
\label{c18}
\end{eqnarray}
Since the right hand side of (\ref{c18}) is not singular, the coefficients $a_n$ and $b_n$ on the left hand side tend to zero for large $n$. The Fourier series on the left hand side can be suitably truncated and the coefficients $a_n$ and $b_n$ evaluated numerically. The FFT algorithms are efficient tools for this purpose.

The complex potential $w_p$ is finally obtained by integrating equation (\ref{c11}):
\be
w_p=\frac{2i\;u_\infty}{\lambda+1}-\frac{Q}{\pi}\log(\lambda+1)+\sum_{n=1}^\infty\frac{a_n-ib_n}{n}\;\lambda^n\;.
\label{c19}
\ee

In general, for geometries defined by the mappings (\ref{c6}) and
(\ref{c7}), the flow does not satisfy the Kutta condition, i.e. it does not separate at the corner. The rotational flow field does exhibit a recirculating bubble, but the separation point is unphysically located downstream from the high curvature corner. To enforce the Kutta condition, a vortex can be added to the flow, as shown in \S 2 for a potential flow: the same considerations regarding its strength and equilibrium hold for a shear flow.

By adding a point vortex to the flow, the complex potential becomes
\be
w_p=\frac{2i\;u_\infty}{\lambda+1}-\frac{Q}{\pi}\log(\lambda+1)+\sum_{n=1}^\infty\frac{a_n-ib_n}{n}\;\lambda^n
+\frac{\gamma}{2i\;\pi}\log\left(\frac{\zeta-\zeta_0}{\zeta-\zeta^\star_0}\right)\;.
\label{c20}
\ee

The complete streamfunction, relevant to the sum of the shear and the potential component of the flow, is
\be
\psi=-\frac{1}{2}\omega\;y^2+\Im(w_p)\;,\label{c20a}
\ee
which provides the following flow velocity:
\be
\dot z^\star=-\omega\;y+\left\lbrack F(\lambda)+\sum_{n=1}^\infty(a_n-ib_n)\lambda^{n-1}\right\rbrack
\frac{d\lambda}{dz}+\left\lbrack\frac{\gamma}{2i\;\pi}\left(\frac{1}{\zeta-\zeta_0}-\frac{1}{\zeta-\zeta_0^\star}
\right)\right\rbrack\frac{d\zeta}{dz}\;.
\label{c21}
\ee
The vortex velocity can be obtained as
\be
\dot z_0^\star=-\omega\;y_0+\lim_{z\rightarrow z_0}\left(\frac{dw_p}{dz}-\frac{\gamma}{2\pi i}\;\frac{1}{z-z_0}\right)
\label{c22}
\ee
and is governed by a Hamiltonian system as in the case of the
potential flow discussed in \S 2:
\be
\dot x_0=\frac{\prz H}{\prz y_0}\;\;\;\;\;\;\;\;\;\;\;\;\;\;\; \dot y_0=-\frac{\prz H}{\prz x_0}\label{c23}
\ee
where the Hamilton function $H$ is:
\be
H=-\frac{1}{2}\;\omega\;y^2+\Im\left\lbrack\frac{2i\;u_\infty}{\lambda+1}-\frac{Q}{\pi}\log(\lambda+1)+\sum_{n=1}^\infty\frac{a_n-ib_n}{n}\;\lambda^n\right\rbrack-\frac{\gamma}{4\pi}\left\lbrack\log \eta_0-\log\left|\frac{dz}{d\zeta}\right|\right\rbrack\;.
\label{c24}
\ee



The strength and location of the vortex is then determined according to equations (\ref{c22}) and (\ref{c23}), imposing the flow at rest on the corner $z_c$ and the vortex at its equilibrium location ($\dot x_0=\dot y_0=0$). Figure 4 shows streamlines and velocity vectors for a separated shear flow for a geometry given by the map (\ref{c6}); the flow field is characterized by $\omega/u_\infty=-10$ and $\gamma/u_\infty=-2.243$. 

The stability of the equilibrium can be inferred from figure 5, which displays the contour levels of the Hamiltonian of the vortex, namely the possible trajectories of the vortex. The equilibrium point, marked with a square, coincide with an elliptic, i.e. stable, fixed point in the phase space. 

\subsection{The unsteady flow}

We examine the Lagrangian transport of passive scalars in the case of unsteady flow. Recalling equations (\ref{a5}) and (\ref{c20a}), the motion is governed by the system of equations:
\be
\dot x=\frac{\prz}{\prz y} \psi(x,y;x_0,y_0);\;\;\;\;\;\;\;\;\;\;\;\;\;\;\; \dot y=-\frac{\prz}{\prz x} \psi(x,y;x_0,y_0)\;,\label{g1}
\ee
where the streamfunction $\psi$ depends on both the particle coordinates ($x,y$) and the vortex location $(x_0,y_0)$. For a vortex located at its equilibrium position ($x_0^E,y_0^E$), the one-degree of freedom Hamiltonian system (\ref{g1}) is autonomous, and hence integrable. In this case, the phase portrait coincides with the streamline pattern for the steady flow as shown in figure 4. The phase space is characterized by a streamline connecting two stagnation points, or according to dynamical system jargon, a heteroclinic orbit connecting two hyperbolic fixed points. Figure 6a shows this streamline in  the physical $z$-plane. Such a feature has great relevance because it is capable of triggering chaotic mixing if the system undergoes a time-dependent perturbation, as shown below.

If the vortex is displaced from its equilibrium location ($x_0^E,y_0^E$) by a small quantity $\varepsilon$, its trajectory follows a closed orbit, as shown in figure 5. Its periodic motion obeys equations of the form:
\be
x_0=x_0^E+f(x_0^E, y_0^E,\varepsilon,t),\;\;\;\;\;\;\;\;\;\;\;y_0=y_0^E+g(x_0^E, y_0^E, \varepsilon,t)\;.\label{g2}
\ee
In this case, system (\ref{g1}) is no longer autonomous, because now the streamfunction (\ref{c20a}) depends on time through the time-dependent vortex coordinates $(x_0(t),y_0(t))$. Formally, we can write
\begin{eqnarray} 
\psi&=&\psi(x,y;x_0^E,y_0^E,\varepsilon)\nonumber\\
 &=&\psi^E(x,y;x_0^E,y_0^E)+\varepsilon\left(\frac{\prz\psi}{\prz x_0}\frac{\prz f}{\prz\varepsilon}
+\frac{\prz\psi}{\prz y_0}\frac{\prz g}{\prz\varepsilon}\right)+O(\varepsilon^2)\;,\label{g3}
\end{eqnarray}
and the governing equations (\ref{g1}) reduce to the time-dependent perturbation of an integrable Hamiltonian system:
\be
\label{g4}
\begin{array}{lcc}
\dot x&=& \displaystyle\frac{\prz\psi^E}{\prz y}+\varepsilon\displaystyle\frac{\prz}{\prz y} \left(\displaystyle\frac{\prz\psi}{\prz x_0}\displaystyle\frac{\prz f}{\prz\varepsilon}
+\displaystyle\frac{\prz\psi}{\prz y_0}\displaystyle\frac{\prz g}{\prz\varepsilon}\right)+O(\varepsilon^2)\nonumber\\  
\dot y&=& -\displaystyle\frac{\prz\psi^E}{\prz x}-\varepsilon\displaystyle\frac{\prz}{\prz x} \left(\displaystyle\frac{\prz\psi}{\prz x_0}\displaystyle\frac{\prz f}{\prz\varepsilon}
+\displaystyle\frac{\prz\psi}{\prz y_0}\displaystyle\frac{\prz g}{\prz\varepsilon}\right)+O(\varepsilon^2)\;.
\end{array}
\ee


As shown in figure 6a, for $\varepsilon=0$ the unstable manifold leaving the hyperbolic fixed point $P^-$ joins smoothly the stable manifold going into the other fixed point $P^+$, and the heteroclinic orbit of the unperturbed system is a single line connecting $P^-$ and $P^+$. However, it is well known that $\varepsilon\ne 0$ dramatically affects the phase portrait in that the time dependence of the system, if not associated to any kind of symmetry, causes non-integrability (Novikov \& Sedov 1979; Aref 1983). The stable and unstable manifolds do not join smoothly but intersect each other transversally forming a paradigmatically chaotic {\em tangle} (e.g. Guckenheimer \& Holmes 1983, Tabor 1989, Wiggins 1992). Since any intersection point maps into another intersection point, the number of transverse intersections is infinite, whereas the distance between intersection points tends to zero as they approach a hyperbolic fixed point.  At the same time, the system preserves the area of the lobes embraced by the arcs of the manifolds clipped between two intersections.
Figure 6b shows a Poincar\'e section of the flow over the snow cornice, enlightening the scenario of the stable and unstable manifold intersecting transversally.   

The chaotic character of a system can be deduced whenever evidence of such transverse crossings is provided. For instance, several authors used the Melnikov technique (Melnikov 1963) to detect analytically the transverse crossings between stable and unstable manifolds for (time-periodically) perturbed integrable systems (e.g. Rom-Kedar, Leonard \& Wiggins 1990, Zannetti \& Franzese 1994, Del Castillo-Negrete 1998 and references therein).

\section{Results}

The representation on the transformed $\zeta$-plane of a steady shearless flow over a step or a snow cornice is similar in many respects to a vortex pair flow, except for the differences due to the Routh correction (\ref{a7}), which affects the vortex velocities. The streamline pattern of a vortex pair flow in the frame moving with the vortices is shown in figure 7, where we can see the two hyperbolic fixed points ($P^-,P^+$) connected by heteroclinic orbits. Rom-Kedar, Leonard \& Wiggins (1990) perturbed the vortex pair by an external periodic strain-rate field, and studied extensively the chaotic dynamics of the resulting Oscillating Vortex Pair flow (OVP). They focused on the dynamics of the lobes enclosed by the perturbed heteroclinic orbits. Rom-Kedar, Leonard \& Wiggins (1990), and Wiggins (1992), provided valuable tools for a quantitative evaluation of the mixing process in terms of the amount of fluid involved and the residence time of the fluid particles in the chaotic region.

The unsteady flows over complex geometries that we described in this paper show the same basic mechanisms of mixing as for the OVP flow.
Besides the fluid permanently entrained by the vortex, and the fluid flowing unperturbed downstream, the scenario is enriched by fluid particles temporarily trapped by the vortex and then detrained, before flowing downstream. This phenomenon, known as {\em transient chaos}, was acknowledged and investigated a few years ago by Pentek, Tel \& Toroczkai (1995) in the context of the advection problem of passive tracers in the velocity field of vortex pairs.


We focus on the phenomenon of heteroclinic chaos as a large scale mechanism of intermittent release, causing intense concentration fluctuation of passive tracers. The entangling of the heteroclinic orbits is a consequence of the unsteadiness of the recirculating region and, in particular, of the oscillatory motion of the reattachment location. This oscillation has been observed in large eddy simulations as well as in two and three-dimensional direct numerical simulation of high Reynolds number flows over a backward-facing step (Driver, Seegmiller \& Marvin 1983, 1987; Le, Moin \& Kim 1997 and references therein). The periodic movement of the reattachment location is caused by the formation and detachment from the step of large-scale coherent vortical structures. The discrete vortex method that we applied to the snow cornice reproduces the same phenomena: figure 3a shows the fluctuations of the total circulation during the transient, due to the periodic detachment of clusters of vortices from the recirculating region. The phenomenon is evident during the transient because no perturbations were introduced in this simulation, but a natural oscillation of the centre of vorticity takes place during the transient, as shown in figure 3b. The transient is characterized by the roll-up of the vortex sheet which is shed into the fluid (see figure 2), consistently with other discrete vortex models of reattached flows (Cortelezzi, Leonard \& Doyle 1994) and DNS of the backward-facing step flow (Le, Moin \& Kim 1997).

As shown below, the unsteady single vortex model  is able to display
the periodic detachment of coherent structures, even though these structures are not vortices but fluid particles trapped into the lobes of perturbed heteroclinic orbits.   

We present the results of two simulations showing the intermittent release of tracers. The first refers to the snow cornice generated by the modified Ringleb transformation (\ref{c6}). We perturbed the flow by displacing the vortex by $\varepsilon=0.005$ and letting it move freely along its periodic orbit. Passive tracers are released at a constant rate from a point upstream of the edge and close to the wall; their time evolution is described by integrating the governing equations (\ref{g1}) by means of a fourth-order Runge-Kutta algorithm. 

The time evolution of the plume is obtained by recording the position of the tracers at fixed time intervals. Also, we record the scalar concentration, averaged over a small control volume, as a function of time. 
Figure 8 shows the distribution of scalars in the field and their
concentration in a sampling volume, marked by a shaded rectangular
area, at two different times. The sampling volume is  downstream. The periodic character of the concentration is demonstrated by its time history, plotted in the upper portion of each frame. The wave form of the concentration was found to be dependent on the location of the monitoring point.

A second example of intermittent release is given by the simulation performed for the geometry obtained by the modified Schwartz-Christoffel mapping (\ref{c7}), i.e. the backward-facing step with smoothed corner. The unsteadiness of the flow is simulated by imposing a random perturbation with zero mean on the position of the vortex in equilibrium. Again, the concentration at a point downstream as a function of time in figure 9 illustrates the intermittent formation of lobes with high density of tracers which are responsible for the concentration fluctuations. Because of the random nature of the perturbation, the concentration is not periodic in time. 

The strongly evident feature of both the simulations is that the perturbation affects the detachment region slightly but the reattachment region quite strongly: the upstream stagnation point moves imperceptibly while the downstream stagnation point has a patently larger movement. This is consistent with the behaviour of real flows around edges, where viscosity effects constrain the separation to be located at the edge, making the separation region rather insensitive to perturbations. From the analytical point of view the two stagnation points in the $\zeta$-plane have oscillations of comparable amplitude, however the mapping onto the physical $z$-plane alters the displacements. The lengths in the $z$-plane are proportional, to first order, to the absolute value of the mapping derivative $|dz/d\zeta|$, whose minimum value is at the edge (for a sharp convex edge $dz/d\zeta=0$). This analytical property provides a reasonable extension of the above mechanism of scalar intermittent release to the geometries characterized by high curvature edges, as well as by separation and reattachment of the flow. 


\section{Conclusions}
This study has examined the effects of the unsteadiness of reattached flows on the large-scale dynamics of dispersion of passive scalars. We considered the two cases of flow past a snow cornice and a backward-facing step, and we used point-vortex models to simulate unsteady separated shear flows. A new technique has been described to determine the streamfunction in a field of constant vorticity. 

A rotational flow over the above geometries reproduces the separation, even though the location of the separation point must be corrected by adding a point vortex to the flow, in order to have separation at the point of maximum curvature. The classical Ringleb mapping for snow cornices, and a Schwartz-Christoffel mapping for step flows have been modified by smoothing the corners. Such modification is necessary to introduce the unsteadiness of the recirculating bubble, in that the separation point is allowed to move around the point of maximum curvature, while having the flow still satisfy an unsteady Kutta condition.

The simulation performed for the snow cornice by a more sophisticated multiple-vortex model detected the oscillation of the center of vorticity, thus showing that the oscillatory motion of the reattachment point is associated with the unsteadiness of the recirculating bubble. The same kind of unsteadiness was imposed in the simulation by the simpler single-vortex model, by perturbing the equilibrium position of the vortex. As the multiple-vortex simulation shows, the flow is characterized by a periodic release of clusters of vortices from the recirculating region behind the crest, in agreement with the direct numerical simulations of the backward-facing step of Le, Moin \& Kim (1997). The same mechanism of releasing vorticity holds for passive tracers. 

Two different perturbations were applied to the system: a periodic, natural perturbation obtained by displacing the vortex from its equilibrium location, and a random motion with zero mean imposed on the vortex in equilibrium. The scalar concentration averaged over small control volumes was calculated as a function of time. For the case of periodic vortex motion, our model detects the intermittent release of blobs of tracers with the same period as the movement of the reattachment location (i.e. the same period as the vortex motion). For the case of random oscillation of the recirculating bubble, the release of coherent structures of tracers was still observed, although the wave form of the concentration is not longer periodic. 

Intermittent dispersion is associated with the phase-space lobe
dynamics (Rom-Kedar, Leonard \& Wiggins 1990, Wiggins 1992), which is
an intrinsically intermittent phenomenon. Since very small
perturbations are sufficient to start the above large-scale
phenomenon, we expect intense concentration fluctuations to appear
downstream whenever a flow undergoes separation and reattachment. 

\bigskip

Much of this work was done while P. Franzese was a postdoctoral
fellow at CSIRO Atmospheric Researh at Aspendale, Australia. We express our appreciation to Dr. Mark Hibberd for his thorough comments,
as well as for his continuous support. We are also grateful to Dr.
Mike Borgas for many helpful discussions and for his encouragements.

\section*{Appendix A}
\subsection*{The Hamiltonian of a vortex}
The Green function $G(\xi,\eta;\xi_0,\eta_0)$ for a flow region $\cal R$ confined by a line $\cal C$ that extends to infinity is defined as
\[G(\xi,\eta;\xi_0,\eta_0)=g(\xi,\eta;\xi_0,\eta_0)+\frac{1}{2\pi}\log\sqrt{(\xi-\xi_0)^2+(\eta-\eta_0)^2}\]
where $g(\xi,\eta;\xi_0,\eta_0)$ is such that $G(\xi,\eta;\xi_0,\eta_0)=0$ on $\cal C$, and is harmonic with respect to $(\xi,\eta)$ on $\cal R$.
Moreover, the reciprocity property of Green functions, i.e. $G(\xi,\eta;\xi_0,\eta_0)=G(\xi_0,\eta_0;\xi,\eta)$, leads to the relationships:
\be
\label{a8a}
\begin{array}{lcl}
\displaystyle \frac{\prz}{\prz \xi_0}g(\xi_0,\eta_0;\xi_0,\eta_0)&=&2\displaystyle\lim_{\zeta\rightarrow \zeta_0}\displaystyle\frac{\prz}{\prz \xi}g(\xi,\eta;\xi_0,\eta_0)\\ 
\displaystyle \frac{\prz}{\prz \eta_0}g(\xi_0,\eta_0;\xi_0,\eta_0)&=&2\displaystyle\lim_{\zeta\rightarrow \zeta_0}\displaystyle\frac{\prz}{\prz \eta}g(\xi,\eta;\xi_0,\eta_0)\;.
\end{array}
\ee
The streamfunction $\psi(\xi,\eta;\xi_0,\eta_0)$ due to a free vortex located in $\zeta=\zeta_0$ must satisfy the boundary condition $\psi=const$ on $\cal C$, and has to be harmonic in the whole region except at $\zeta_0$, where it behaves as $(\gamma/2\pi)\log\sqrt{(\xi-\xi_0)^2+(\eta-\eta_0)^2}$.
Therefore, the streamfunction $\psi$ is related to the Green function by:
\be
\psi(\xi,\eta;\xi_0,\eta_0)=\psi_1(\xi,\eta)+\gamma G(\xi,\eta;\xi_0,\eta_0)\label{a7a}
\ee
where $\psi_1$ is the streamfunction of other flows possibly superimposed on the vortex.

A vortex does not induce velocity on itself. Therefore, it moves as a particle in the flow field from which the vortex induction has been removed. We can calculate the complex conjugate velocity of a vortex as: 
\be
\dot \zeta_0^\star=\dot \xi_0-i\;\dot \eta_0=\lim_{\zeta\rightarrow \zeta_0}\left(\frac{dw}{d\zeta}-\frac{\gamma}{2\pi i}\frac{1}{\zeta-\zeta_0}\right)
=\left(\frac{\prz\psi_1}{\prz\eta}+i\frac{\prz\psi_1}{\prz\xi}\right)_{\zeta=\zeta_0}+\gamma \lim_{\zeta\rightarrow \zeta_0}\left(\frac{\prz g}{\prz \eta}+i\frac{\prz g}{\prz\xi}\right)
\label{a8}
\ee
where $w$ is the complex potential. According to equations (\ref{a8a}), the velocity of a vortex can be expressed as:
\be
\dot\zeta_0=\frac{\prz H'}{\prz\eta_0}+i\frac{\prz H'}{\prz\xi_0}
\label{a8aa}
\ee
where the Hamiltonian $H'$ is given by:
\be
H'=\psi_1+\frac{\gamma}{2}g\;.
\label{a8b}
\ee
For instance, in our case the curve $\cal C$ and the flow region $\cal R$ are the real axis and the positive imaginary half-plane respectively, and the flow is the superimposition of a uniform flow and a vortex. Therefore the vortex Hamiltonian $H'$ (\ref{a8b}) becomes:
\be
H'=q_\infty\eta_0+\frac{\gamma}{4\pi}\log{\eta_0}\;.
\label{a8cb}
\ee

\subsection*{The Routh rule}
When the motion of a vortex is conformally mapped onto the physical $z$-plane, its Hamiltonian has to be corrected according to the so-called Routh rule (Routh 1881). The following is a brief account of its derivation. 

The complex conjugate velocity of a vortex in the $z$-plane is
\be
\dot z_0^\star=\dot x_0-i\;\dot y_0=\lim_{z\rightarrow z_0}\left(\frac{dw}{dz}-\frac{\gamma}{2\pi i}\;\frac{1}{z-z_0}\right)\;.
\label{a11}
\ee
This can be written as:
\be
\dot z_0^\star=\lim_{\zeta\rightarrow \zeta_0}\left\lbrack\left(\frac{dw}{d\zeta}-\frac{\gamma}{2\pi i}\frac{1}{\zeta-\zeta_0}\right)   \frac{d\zeta}{dz}+\frac{\gamma}{2\pi i}\left(\frac{1}{(\zeta-\zeta_0)\displaystyle\frac{dz}{d\zeta}}-\frac{1}{z-z_0}\right)\right\rbrack\label{a12}
\ee
that is
\be
\dot z_0^\star=\dot\zeta_0^\star\frac{d\zeta_0}{dz_0}-\frac{\gamma}{2\pi i}\frac{\displaystyle\frac{d^2 z_0}{d\zeta^2_0}}{2\left(\displaystyle\frac{d z_0}{d\zeta_0}\right)^2}
=\dot\zeta_0^\star\frac{d\zeta_0}{dz_0}-\frac{\gamma}{4\pi i}\frac{d}{dz_0}\log\left(\frac{dz_0}{d\zeta_0}\right)\;.
\label{a13}
\ee
Recalling equations (\ref{a8aa}), it follows that
\begin{eqnarray}
\dot z_0^\star&=&\frac{\prz H'}{\prz\xi_0}\frac{\prz\xi_0}{\prz y_0}+\frac{\prz H'}{\prz\eta_0}\frac{\prz\eta_0}{\prz y_0}
+i\left(\frac{\prz H'}{\prz\xi_0}\frac{\prz\xi_0}{\prz x_0}+\frac{\prz H'}{\prz\eta_0}\frac{\prz\eta_0}{\prz x_0}\right)+
\frac{\gamma}{4\pi}\left(\frac{\prz}{\prz y_0}\log\left|\frac{dz_0}{d\zeta_0}\right|+i\frac{\prz}{\prz x_0}\log\left|\frac{dz_0}{d\zeta_0}\right| \right)\nonumber\\
 &=&\left(\frac{\prz}{\prz y_0}+i\frac{\prz}{\prz x_0}\right)\left(H'+\frac{\gamma}{4\pi}\log\left|\frac{dz_0}{d\zeta_0}\right|\right)
\label{a14}
\end{eqnarray}
and finally
\be
H=H'+\frac{\gamma}{4\pi}\log\left|\frac{d z_0}{d \zeta_0}\right|\,.\label{a7b}
\ee
The relationship (\ref{a8b}) between a vortex Hamiltonian and a Green function for a bounded simply connected region was first given by Masotti (1931) (but see also Caldonazzo 1931 and Pelosi 1926). Successively, Lin (1941) extended the Masotti theory to multiply connected domains containing more than one vortex. Surprisingly, Ringleb (1961) seems to have been unaware of the previous studies, as he deduced independently the Routh rule. However, he considered that it was only applicable for mappings based on many-valued functions.

\newpage
\begin{center}
\section*{\normalsize REFERENCES}
\end{center}

\hangindent=4truemm\hangafter=1 
\noindent{Aref, H. 1983 Integrable, Chaotic, and Turbulent Vortex
Motion in Two Dimensional Flows. {\em Ann. Rev. Fluid Mech.} {\bf 15}, 345-389.}

\hangindent=4truemm\hangafter=1 
\noindent{Bradshaw, P. \& Wong, F. W. F. 1972 The reattachment and relaxation of a turbulent shear flow. {\em J. Fluid Mech.} {\bf 52}, 113-135.}

\hangindent=4truemm\hangafter=1 
\noindent{Caldonazzo, B. 1931 Sui moti liquidi piani con un vortice libero. {\em Rend. Circ. Matem. Palermo} {\bf LV}, 369-394.}

\hangindent=4truemm\hangafter=1 
\noindent{del Castillo-Negrete, D. 1998 Asymmetric transport and non-Gaussian statistics of passive scalars in vortices in shear. {\em Phys. of Fluids} {\bf 10}, 576-594.}

\hangindent=4truemm\hangafter=1 
\noindent{Ceschini, E. 1993 {\em Interferenza vortice-parete}. Laurea thesis, DIASP, Politecnico di Torino.}

\hangindent=4truemm\hangafter=1 
\noindent{Clements, R. R. 1973 An inviscid model of two-dimensional vortex shedding. {\em J. Fluid Mech.} {\bf 57}, 321-336.}

\hangindent=4truemm\hangafter=1 
\noindent{Cortelezzi, L., Leonard, A. \& Doyle, J. C. 1994 An example of active circulation control of the unsteady separated flow past a semi-infinite plate. {\em J. Fluid Mech.} {\bf 260}, 127-154.}

\hangindent=4truemm\hangafter=1 
\noindent{Driver, D. M., Seegmiller, H. L. \& Marvin, J. 1983 Unsteady behaviour of a reattaching shear layer. {\em AIAA Paper} 83/1712.}

\hangindent=4truemm\hangafter=1 
\noindent{Driver, D. M., Seegmiller, H. L. \& Marvin, J. 1987 Time-dependent behaviour of a reattaching shear layer. {\em AIAA J.} {\bf 25}, 914-919.}

\hangindent=4truemm\hangafter=1 
\noindent{Ferlauto, M. 1996 {\em Studio e controllo di strutture vorticose di parete.} Ph.D. thesis, DIASP, Politecnico di Milano.}

\hangindent=4truemm\hangafter=1 
\noindent{Guckenheimer, J. \& Holmes P. 1983 {\em Non-linear oscillations, dynamical systems and bifurcation of vector fields}. Springer-Verlag.}

\hangindent=4truemm\hangafter=1 
\noindent{Henrici, P. 1974 {\em Applied and Computational Complex Analysis}, John Wiley \& Sons.}

\hangindent=4truemm\hangafter=1 
\noindent{Hosker, R. P. 1984 Flow and diffusion near obstacles. {\em Atmospheric science and power production}, U.S. Dept. of Energy, 241-326.}

\hangindent=4truemm\hangafter=1 
\noindent{Hunt, J. C. R., Leibovich, S. \& Richards, K. J. 1988 Turbulent shear flows over low hills. {\em Q. J. R. Meteorol. Soc.} {\bf 114}, 1435-1470.}

\hangindent=4truemm\hangafter=1 
\noindent{Le, H., Moin, P. \& Kim, J. 1997 Direct numerical simulation of turbulent flow over a backward-facing step. {\em J. Fluid Mech.} {\bf 330}, 349-374.}
 
\hangindent=4truemm\hangafter=1 
\noindent{Lin, C. C. 1941 On the Motion of Vortices in Two
Dimensions - I. Existence of the Kirchhoff-Routh Function, {\em Proc.
N.A.S.} {\bf 27}, 570-575.}

\hangindent=4truemm\hangafter=1 
\noindent{Masotti, A. 1931 Sulla funzione preliminare di Green per un'area piana. {\em Atti Pont. Accad. Sci. Nuovi Lincei} {\bf 84}, 209-216. Sul moto di un vortice rettilineo. {\em Ibidem}, 235-245. Sopra una proprieta' energetica di un vortice rettilineo. {\em Ibidem}, 464-467. Sopra una relazione di reciprocita' nella idrodinamica. {\em Ibidem}, 468-473. Sulle azioni dinamiche dovute ad un vortice rettilineo. {\em Ibidem}, 623-631.}

\hangindent=4truemm\hangafter=1 
\noindent{Melnikov, V. K. 1963 On the stability of the center for time periodic perturbations. {\em Trans. Moscow Math. Soc.} {\bf 12}, no. 1.}

\hangindent=4truemm\hangafter=1 
\noindent{Nehari, Z. 1975 {\em Conformal Mapping}, Dover Publications Inc.}

\hangindent=4truemm\hangafter=1 
\noindent{Novikov, E. A. \& Sedov Y. B. 1979 Stochastization
of Vortices. {\em JETP Lett.} {\bf 29}, no. 12, 677-679.}

\hangindent=4truemm\hangafter=1 
\noindent{Pelosi, L. 1926 Un'applicazione idrodinamica della funzione di Green. {\em Atti Reale Accad. Sci. Torino} {\bf LXI}, 569-583.}

\hangindent=4truemm\hangafter=1 
\noindent{Pentek, A., Tel, T. \& Toroczkai, Z. 1995 Chaotic advection in the velocity field of leapfrogging vortex pairs. {\em J. Phys. A Math. Gen.} {\bf 28}, 2191-2216.}

\hangindent=4truemm\hangafter=1 
\noindent{Ringleb, F. O. 1961 Separation control by trapped vortices, in {\em Boundary Layer and Flow Control}, edited by G. V. Lachmann, Vol. 1, Pergamon Press, 265-294.}

\hangindent=4truemm\hangafter=1 
\noindent{Rom-Kedar, V., Leonard, A. \& Wiggins, S. 1990 An analytical study of transport, mixing and chaos in an unsteady vortical flow. {\em J. Fluid Mech.} {\bf 214}, 347-394.}

\hangindent=4truemm\hangafter=1 
\noindent{Routh, E. J. 1881 Some applications of conjugate functions. {\em Proc. Lond. Math. Soc.} {\bf 12}, 83.}

\hangindent=4truemm\hangafter=1 
\noindent{Tabor, M. 1989 {\em Chaos and integrability in non linear dynamics}, John Wiley \& Sons.}
 
\hangindent=4truemm\hangafter=1 
\noindent{Tsien, H. S. 1943 Symmetrical Joukowsky Airfoils in shear flow. {\em Q. Appl. Math.} {\bf 1}, 130-148.}

\hangindent=4truemm\hangafter=1 
\noindent{Wiggins, S. 1992 {\em Chaotic Transport in Dynamical Systems}. Springer-Verlag.}

\hangindent=4truemm\hangafter=1 
\noindent{Zannetti, L. \& Franzese, P. 1994 The non-integrability of the restricted problem of two vortices in closed domains. {\em Physica D} {\bf 76}, 99-109.}

\clearpage

\begin{center}
\section*{\normalsize \bf FIGURE CAPTIONS}
\end{center}

Fig. 1: (a) The snow cornice obtained by the Ringleb mapping; (b) the backward-facing step obtained by the Schwartz-Christoffel mapping.

Fig. 2: Simulation of the flow past a snow cornice by a multiple vortex method. The solid circles represent the vortices shed throughout the transient. The vortices strengths are proportional to the symbol sizes.

Fig. 3: (a) Total circulation for the multiple vortex method (solid
line) and circulation for the single vortex model (dotted line), as a function of time; (b) distance of the centroid for the multiple vortex method from the equilibrium point of a single vortex. 

Fig. 4: Streamlines and velocity vectors for a separated shear flow for the blunt snow cornice.

Fig. 5: Pattern of the possible trajectories of the vortex. The equilibrium point is marked with a square. 

Fig. 6: (a) The separatrix streamline connecting two fixed points ($P^-, P^+$) in the unperturbed case; (b) as a consequence of the perturbation, the streamline splits in two branches which intersect each other: unstable manifold (bold line) and stable manifold (thin line).

Fig. 7: Streamline pattern of a vortex pair flow in the frame of reference moving with the vortices.

Fig. 8: Distribution of scalars in the field behind a snow cornice and concentration time series sampled in the volume marked by the shaded rectangular area downstream, at two different times. The concentration time series is plotted in the upper portion of each frame. The system is periodically perturbed by the vortex autonomously oscillating around its equilibrium location.

Fig. 9: Distribution of scalars in the field behind a backward facing
step with smoothed corner and concentration time series sampled in the volume marked by the shaded rectangular area downstream, at four different times. The concentration time series is plotted in the upper portion of each frame. A random perturbation is imposed on the vortex equilibrium location. 

\end{document}